\documentstyle[12pt,aaspp4]{article} 
\begin{document}
\title{Complex extended line emission in the cD galaxy in Abell 2390} 
 
\author{J.B. Hutchings\altaffilmark{1}} 
\affil{Dominion Astrophysical Observatory\\Herzberg Institute of Astrophysics,
National Research Council of 
Canada\\ 5071 W. Saanich Rd., Victoria, B.C. V8X 4M6, Canada} 
 
\author{M.L. Balogh\altaffilmark{2,3}}
\affil{Dept of Physics and Astronomy, University of Victoria\\
P.O. Box 3055, Victoria, B.C. V8W 3P6, Canada}

\authoremail{john.hutchings@hia.nrc.ca}

\altaffiltext{1}{Based on observations with the NASA/ESA \it Hubble Space 
Telescope,
\rm obtained at the Space Telescope Science Institute, which is operated
by AURA Inc under NASA contract NAS5-26555}
\altaffiltext{2}{Visiting Astronomer, Canada--France--Hawaii Telescope, which is 
operated by the National Research Council of Canada, le Centre Nationale de la Recherche Scientifique, and the University of Hawaii.}
\altaffiltext{3}{Present address: Department of Physics, University of Durham, South Road, Durham, England DH1 3LE}

\begin{abstract}

  This paper reports maps of the cD galaxy in the rich z=0.23 cluster Abell 2390 at UV, [OII], L$\alpha$ and H$\alpha$
wavelengths.  Spatially resolved UV and optical spectra were obtained with STIS on
the {\it Hubble Space Telescope}; the 2\arcsec\ wide slit
was aligned close to the long axis and blue lane of the galaxy and includes all 
the inner bright features seen in WFPC2 images. The L$\alpha$ is seen in emission 
from several bright knots and over an extended region from 4\arcsec\ NW 
to 2\arcsec\ SE of the nucleus. Three of these knots
have detected UV continuum as well.  H$\alpha$ images were obtained with {\it OSIS} at
CFHT; both H$\alpha$ and [O II] show extended emission that roughly trace L$\alpha$,
primarily to the NW.  Notable differences between the spatial
distributions of H$\alpha$, L$\alpha$ and [O II] emission
and the blue knots in the galaxy may be the result of inhomogeneous dust
extinction, or variations in ionisation.
The L$\alpha$ emission velocities depend the uncertain identification with features in the
undispersed images; there is strong evidence for resolved
emission knots to have large velocities, $\gtrsim$1000 km/s, indicative of infall.
The L$\alpha$ spectrum shows a sharp cutoff which may be due to absorption; we 
entertain the possibility that this edge is due to an shell of absorbing gas, outflowing
at $\sim$5000 km/s.

\end{abstract}

\section{Introduction}
The central galaxies of rich clusters often differ remarkably
from other cluster ellipticals in their morphological and spectroscopic properties;
in particular, a large fraction ($\sim 40$\%) show
strong nebular emission lines and an excess ultraviolet/blue continuum (e.g., Johnstone, Fabian
\& Nulsen 1987; Heckman et al. 1989; McNamara \& O'Connell 1993; Crawford et al. 1999).
These features are more common in clusters selected from X-ray samples, and especially
in those which are known to have strong cooling flows.  The optical line emission is generally
very concentrated in a central region of only 5--10 kpc (e.g. Heckman et al. 1989; Crawford et al. 1999),
sometimes with filaments extending beyond the stellar continuum, 
$\sim 20$ kpc into the intracluster medium (e.g. Cowie et al. 1983; Romanishin \& 
Hintzen 1988; Crawford \& Fabian 1992).  The cooling flow gas itself extends to much larger
scales ($\sim$100 kpc).  

The origin of the nebular emission in cooling flow clusters is uncertain, and is likely
not the same in all clusters.  
Generally, the line luminosity is too strong to arise directly from the
cooling gas.  Heckman et al. (1989) showed that the spatial variation of emission line
ratios generally corresponds better with shock models, rather than photoionization models.
Other authors (e.g. Filippenko \& Terlevich 1992, Allen 1995) claim the line ratios can
be produced by very massive O-stars, which may form in the collision of cold gas clouds
within the cooling flow.
The star formation rates determined from this nebular emission, assuming the
gas is photoionised by massive stars, are generally 10--100 $M_\odot$ yr$^{-1}$ (e.g. Johnstone et al. 1987;
McNamara \& O'Connell 1993; Allen 1995; Crawford et al. 1999), which are at least a factor of 10 less than the mass inflows
implied by the cooling flows.  

The luminous emission line nebulae in these unusual galaxies are often
asymmetrically distributed about the galactic
nucleus (Crawford \& Fabian 1992).  McNamara \& O'Connell (1993) observe strong colour
structure in the blue spectra of two such galaxies which have smooth I-band isophotes.
In some cases, this structure is seen to correlate with features at radio wavelengths
(e.g. Heckman et al. 1989; Edge et al. 1999) which suggest ionisation of the gas
is at least partly due to a central radio source.  The velocity structure of the nebulae
also tends to be complex.  Heckman et al. (1989) found that the velocities of H$\alpha$
nebulae in nine central galaxies are very disordered (uncorrelated with position), with
velocities of $\lesssim 200$ km/s.  They claim this is consistent with a scenario in 
which the gas is clumpy and infalling (in a cooling flow), such that the ``random'' fluctuations
in velocity are due to clumps observed in both the foreground and background of the centre.
Occasionally much higher velocities are observed; Johnstone \& Fabian (1995) observe Ly$\alpha$
emission and absorption velocities at 1000 km/s and greater, which are likely also due to nebulae
infalling along the line of sight (Haschick, Crane \& van der Hulst 1982). 

Recent observations have detected large amounts of dust in some central galaxies,
often amounting to $E_{B-V}\gtrsim 0.5$ (e.g. Hu 1992; 
Allen et al. 1995; McNamara et al. 1996; Pinkney et al. 1996; Crawford et al. 1999).  
The presence of such large amounts of dust may itself require a high
star formation rate to replenish the supply destroyed by sputtering (e.g. Draine \& Salpeter 
1979).
Neglecting this dust when analysing emission line fluxes can
lead to an underestimation of the implied star formation rates by an order of magnitude.

The cluster Abell 2390 is host to a particularly strong cooling flow, with an
inflow rate of $\sim 800 M_\odot$ yr$^{-1}$ derived from {\it ROSAT} observations
(Pierre et al. 1996).  The galaxy populations in this cluster have been studied in some detail by the Canadian
Network for Observational Cosmology (CNOC1)
consortium, using images and spectra from the Canada France Hawaii Telescope
(Yee \it et al \rm 1996, Abraham \it et al \rm 1996); evidence for infall and
strong population gradients throughout the cluster is seen. The central bright cD galaxy
of the cluster has considerable size and structure and was studied using
the CNOC1 database by Davidge and Grinder (1995). It has a lumpy
morphology at resolutions of a few arcseconds or better, extremely blue ($g-r$)
and ($U-B$) colour (Smail et al. 1998) and a predominantly young stellar population, with
strong [O II] emission (rest frame equivalent width of
110 $\pm$ 2 \AA). Davidge and Grinder
note that, in this respect, this galaxy is similar to the cD galaxies of CNOC1 clusters 0839+29 and
1455+22, and different from other cDs in that survey. 
This galaxy has also been studied in detail
at submillimetre, radio, infrared and optical wavelengths by Edge et al. (1999); in particular,
these authors note the presence of a strong blue lane (or ``cone'') in the archival WFPC2 observations
which is oriented orthogonally to the 4.89 GHz radio map, suggesting
that the gas in this cone is ionised by a strong nuclear source.
Edge et al. and Lemonon et al. (1998) claim that there is some evidence for dust in this galaxy, though
this is strongly dependent on the uncertain nature of the ionisation source.

We have obtained {\it HST} STIS observations of the cD galaxy in A2390 
with a 2\arcsec\ slit, which provides spatial and velocity information about the
Ly$\alpha$ and [OII] emission lines, and the UV continuum.  We compare these
data
with CFHT H$\alpha$ observations obtained as part of a larger study (Balogh \& Morris 1999).
With this data, we map the structure of nebular emission within $\sim 10$ kpc of
the galaxy centre.  We present our observations in \S\ref{sec-obs}, and our results on
the spatial and velocity structure of the emission line gas in \S\ref{sec-results}.  In
\S\ref{sec-discuss} we discuss some of the implications of these observations.  Our
results are summarized in \S\ref{sec-summary}.

\section{Observations}\label{sec-obs}
\subsection{STIS Observations}\label{sec-Lyalpha}
{\it HST} STIS observations were obtained 
with the 2\arcsec\ wide slit aligned at  $\sim 25^\circ$ to the direction of the blue lane
seen in the WFPC2 images (\S\ref{sec-wfpc}).  Standard reduction procedures (flats and wavelength
calibration) were followed. 
The slit direction was chosen to include another central cluster galaxy, and 
two of the gravitational arcs near
the cluster centre, for additional information. Spectra were taken at both optical and UV wavelengths,
as summarized in Table 1.
The blue G430L spectrum has resolved
[O II] and H$\gamma$ emission, while the G750L exposure is too short to
be useful, though it does show H$\alpha$ emission and continuum.  L$\alpha$
emission is clearly resolved in the FUV MAMA spectra.
The very wide slit ensured that all bright central features
in the galaxy were included, and the resulting spectra are essentially slitless,
so that spectral feature positions depend on their physical location as
well as their radial velocities. 
The STIS observation first centred on a nearby bright star, and then moved
to the galaxy by accurate blind offset. Since the galaxy is extended, its
coordinates may not be exactly those of the nucleus, and there is some
uncertainty  (perhaps $\sim$0.1") in the exact location of the 2" wide slit.

\subsection{H$\alpha$ Imaging}
Images of the cD galaxy in H$\alpha$+N[{\small\rm II}]\footnote{
Hereafter, we refer to this as H$\alpha$ alone; corrections for N[{\small\rm II}] are
made when necessary.} light were taken with {\it OSIS} on the CFHT,
with a specially designed interference filter (Balogh 1999, Balogh \& Morris 1999). 
A continuum image, formed
by combining narrow band images redward and blueward of H$\alpha$, was subtracted
from the on-line image, after matching the PSF (0\farcs8), to produce the final image.  Full details of
the reduction procedure are given in Balogh (1999).
The H$\alpha$ image 
was aligned with the WFPC2 images by matching the
H$\alpha$ continuum image with the WFPC2 F814W image, and this
should be good to about 0\farcs1.  
H$\alpha$ fluxes could not be accurately determined, since the nights were not photometric, but 
equivalent widths are reliably measured.
The H$\alpha$ filter used was quite wide, 324 \AA\ FWHM, so H$\alpha$ emitted
at $v\lesssim 6000$ km s$^{-1}$ relative to the cD galaxy will be detected.

\subsection{Comparison with Archival WFPC2 Images}\label{sec-wfpc}
We have obtained images of the cD galaxy in Abell 2390 from the HST archive;
it has been observed with the WFPC2 with the F555W and F814W filters.  
Figure~\ref{wfpc}
shows all of the images with the same spatial scale, and also the width of the
slit, the length of which is vertical in the orientation of the diagram.
The first panel shows the WFPC2 F555W image, and the third panel presents
the F814W/F555W ratio, thus showing the colour distribution
(light shades correspond to blue colours).  
The horizontal, solid line shown in the F555W image represents the size and
approximate location of the 2\arcsec\ slit used to obtain the STIS spectra.  
At the cluster redshift of $z=0.23$,
1\arcsec\ corresponds to 3.4 kpc\footnote{We assume a cosmology with H$_\circ$=75 km s$^{-1}$ Mpc$^{-1}$, $\Omega_\circ$=0.3 and $\Omega_\Lambda=0.7$.}.
We consider the redder
knot that lies in the centre of the general galaxy light to be the nucleus; this feature
is surrounded by red material.  There is also a red, resolved knot, 23\arcsec\ E of S from
the galaxy that is a separate
small galaxy of red colour
(number 956 in Yee {\it et al} 1996).  It is a normal-looking elliptical with an old stellar
population, and the spectra and H$\alpha$ images do not reveal any spatially varying properties 
or line emission. It is too far away from the cD galaxy to appear in the UV
spectral image, which covers only 25\arcsec. There is a faint signal at the
position of the bright arc about 8\arcsec\ S of the cD, which is flat and
without strong emission in the observed blue wavelength range. This is
consistent
with a very blue stellar population, at redshift larger than 0.23. There
is no significant signal at the position of a larger red arc about 14\arcsec\
from the cD galaxy. We do not discuss these other objects further.

\section{Results}\label{sec-results}
\subsection{Spatial Structure}
The most prominent blue features seen in the WFPC2 images of the cD galaxy 
in Figure \ref{wfpc} are the resolved knots below the nucleus, just above
the nucleus, and the linear feature above the nucleus. These all lie roughly
along a line.  Edge et al (1999) interpret the blue lane as a double ``conical'' feature,
similar to that seen in Seyfert galaxies, and suggest that this material may
be ionised by the strong nuclear source.

   The UV spectrum shown in the right hand panel of Figure~\ref{wfpc}
(and, spatially compressed, in Figure~\ref{uvspec}) shows a remarkable
L$\alpha$ complex. The direct and dispersed images are
lined up exactly in the spatial direction along the slit.
The 2\arcsec\ slit covers the brightest, central region ($\sim$7 kpc)
of the galaxy and its various
bright knots; the continuum profile of the halo itself has an effective radius
of about 10\arcsec.   There is L$\alpha$ emission that extends over the whole length 
of the inner galaxy, and is considerably more extended below (NW) the nucleus, 
beyond the bright blue knots. The brightest L$\alpha$ knots are the nucleus itself
and the knots above and below it in our image.

There is UV continuum from three of the knots, including
one fainter blue knot below the nucleus. However, continuum light
from the nucleus itself is not detected. The CCD blue spectral image
shows that there is continuum at the nucleus and the bright knots above
it in our diagram, but not from the knot below which is bright in the UV.
The [O II] line emission is brightest at the nucleus but is seen clearly
at the two L$\alpha$ knots as well. The blue lane does not correspond with
either line or continuum flux in our data, so presumably arises from a
combination of lines and continuum in the WFPC2 passbands.

  There is a prominent, short wavelength edge to the L$\alpha$ 
emission. This is shown more clearly in Figure~\ref{uvspec}, which is 
compressed 
along the spatial direction to make this edge and its curvature more apparent.
Where the edge crosses the lower continuum region, there is some evidence for
absorption over an interval shortward of the edge; this can be seen in 
spectral profile shown in Figure \ref{profiles}.  However, the shape of the continuum is not well defined,
which makes this determination uncertain.  

As noted above, in contrast to the L$\alpha$ image, the nucleus is the 
brightest feature in the dispersed [O II] image.  Furthermore, 
no evidence for the sharp L$\alpha$ edge is seen
in the [O II] profile, as shown in Figure \ref{profiles}. 
We note that the [O II] profile we show is predominantly that of the nuclear region:
the contribution from the L$\alpha$ bright knot is small and does not affect
the shortward side of the profile we show. The presence of [O II] emission
at velocities more negative than those at L$\alpha$ suggests that
the L$\alpha$ edge  \it may \rm be due to absorption, since the [O II] 
doublet (velocity separation $\sim$220 km s$^{-1}$) is from forbidden
transitions.
However, the [OII] emission signal is too weak to enable us to draw strong
conclusions. The velocity structure of the L$\alpha$ edge and of the various
emission features is discussed in more detail in \S\ref{sec-velocities}.

The image of the cD galaxy in H$\alpha$ light is shown in the left hand
panel of Figure \ref{wfpc}.  
The nuclear regions of the galaxy, and at least two regions extending several
arcseconds below (NW) the nucleus, are clearly detected. 
The photometric zero point of the H$\alpha$ image is uncertain by at least $\sim 0.3$ mag,
as observing conditions were poor; the rest frame equivalent width
line (which is independent of this zero point) is 150$\pm7$ \AA\ within a 5\arcsec\ diameter aperture.

The nuclear region is extended towards the knot above (in Figure \ref{wfpc}) 
the nucleus, but does not show a peak
at the same position.  The first H$\alpha$ emission region below the nucleus
corresponds with the faint inner knots in the WFPC2 image, and extends 
towards the brighter blue knot below. 
The absence of detected H$\alpha$ emission to the SE of the
nucleus, along the clearly visible blue lane, is notable.  However, the H$\alpha$ data are not
very sensitive; the limiting H$\alpha$ magnitude is
$m_{AB}=21.7$ (for a 2$\sigma$ point source detection), which corresponds to a flux limit of
1.6$\times10^{41}$ ergs/s.  
Weaker emission  may be present along the blue lane;
clearly, however, the strongest emission originates from the NW
of the nucleus.

The L$\alpha$ bright knot emission distribution appears to correspond more closely
with the features in the HST continuum images, rather than the
H$\alpha$ image.  However, this comparison is complicated by the very different
spatial resolutions, and the faintest emission boundaries do 
resemble those of H$\alpha$ if we assume that the slit edge cuts off the leftmost
part of the L$\alpha$ emission.
The H$\alpha$ emission peak farther to
the NW extends towards the two fainter blue knots farthest from the nucleus, 
but does not correspond in detail with their positions. Plots of the
row-averaged light from the H$\alpha$, F555W, and L$\alpha$ images (not shown)
confirm that they differ significantly. In particular, the L$\alpha$
emission is strong at the blue knots with UV continuum, while H$\alpha$
is not; there are other places where the reverse is true.  Similar effects
are seen in nearby starburst galaxies, which Conselice et al. (1999) claim
are consistent with a picket fence dust distribution.  However, it is still 
unclear whether, in this case, the observed structure is mostly due to differences in
extinction, ionisation, or sensitivity.

  We note finally that we have a VLA C-configuration 1.4GHz map of the galaxy that shows
a weak extension in the general direction of the NW knots. Edge et al (1999)
claim that the A and B configuration reveal a smaller (0.3\arcsec) extension at
5 GHz, possibly normal to the blue lane. Deeper, high resolution radio maps at
several frequencies would be of interest in understanding whether there are
jets along the blue lanes or not.

\subsection{Velocity Structure}\label{sec-velocities}
Because of the wide slit, there is ambiguity between spatial and velocity
structure in the STIS spectra which plagues our analysis of the nebular velocities.
The geocoronal emission produces an emission line
that is 50\AA\ wide, from uniformly distributed emission at zero radial velocity.
A similarly distributed L$\alpha$ emission wider than the 2\arcsec\ wide slit around
the galaxy would produce the same uniformly illuminated spectral feature
from 1465-1526\AA. Line emission from spatially resolved knots within the
slit will be shifted by the combination of radial velocity and position within
the slit, broadened by the intrinsic motions and spectroscopic resolution.

To measure the velocities of resolved knots, which one can hope to locate
spatially from continuum images,
we matched the dispersed and undispersed image features in the dispersion
direction and noted the relative offsets for as many recognisable components 
as possible. This removes the spatial shifts from the dispersed image
before measuring velocity shifts, a technique that has been used and described
e.g. in Hutchings  \it et al \rm (1998) for NGC 4151 slitless data. The
quantitative measures were obtained by superposing contour plots of knots 
in the dispersed and undispersed images. This does not provide an absolute
scale, so the zero point was set by identifying the nucleus as
described above and referring all other velocity shifts to it. The nuclear
velocity is assumed to have the ground-based redshift 0.2301 (Abraham et al 1996)
derived from a narrow slit spectrum of the central galaxy. (Yee \it et al \rm 1996)
published a redshift of 0.23024. For the purpose of this paper,
the difference, which amounts to some 40 km s$^{-1}$ is not significant, and
is within the quoted errors of Yee et al.)

This process assumes that L$\alpha$ emission, aside from the brightest knots,
arises at the peak of the F555W image flux - in particular, along the blue lanes on either
side of the nucleus. We regard this as reasonable given the correspondence in
overall flux distributions along the galaxy. However, this can only be confirmed
by narrower slit observations. If, for example, the L$\alpha$ emission 
arises along the faint extensions in the H$\alpha$ image to the upper left and
lower right, rather than the continuum flux along the blue lanes, then the
emission velocities will be quite different. 
Figure~\ref{vel} shows the emission knot velocities based on the blue
continuum image and also the H$\alpha$ image. 
The values derived from the blue continuum image form
a rough S-shape, symmetrical about a point 0\farcs2 below the nucleus, about
mid-way between the outer UV-continuum knots. 
The central region has a `rotation' curve of amplitude about 400 
km s$^{-1}$. Further out, this curve reverses, reaching a full 
amplitude more than 7000 km s$^{-1}$ across the outer regions.  Weaker evidence
for such high velocities are also present in the [OII] spectra.  The velocities
we derive if the L$\alpha$ arises in the H$\alpha$ regions (dots in
Figure~\ref{vel}) are also high. Note that the dots in Figure~\ref{vel} at
positions $<-1$\arcsec\
indicate double-valued velocities, since the H$\alpha$ emission splits and
forms two ``tails'' in this region. This significantly reduces the velocities 
in the SW, where identification of L$\alpha$ with the blue lane is most in doubt.
However, the measured velocities corresponding to the knots are still quite high, and 
probably arise in in- or out-flowing material. 

The curved L$\alpha$ edge is a particularly striking feature.
It corresponds closely to the position of the edge of
the slit, and this could give rise to this
feature if there is extended L$\alpha$ emission across much of the galaxy,
with weak velocity structure to produce the curvature.
It is difficult to know the slit position exactly as the geocoronal line
is very strong and has broad wings. In addition, there is a zero point
shift for the whole velocity scale, derived from a different narrow
slit, and there is some uncertainty in the precise redshift from an
extended source like this galaxy. 
However, if the L$\alpha$ edge is caused by the slit position, the curvature
and other morphology seen in the dispersed image likely reflects
primarily the spatial structure of the ionised gas;
the overall shape of the faintest emission is similar for undispersed
H$\alpha$ and dispersed L$\alpha$, suggesting that this may be the case.

However, there are some reasons to believe that the sharp edge corresponds instead
to a real absorption feature:

1.  The line emission is not evenly distributed, but generally peaks 
near the shortward edge and falls off slowly 
towards longer wavelengths. The emission image has no redward edge 
although we
might expect emission to cover that part of the galaxy, especially at
the top of our diagrams, where the galaxy tilts to that side.
If this is spatial structure, we have the
unlikely situation where emission arises preferentially
toward the slit.  

2. The shortward edge is not straight, but has smooth curvature over 
a large distance.  Since it appears to curve in a ``C'' shape,
rather than the ``S'' shape characteristic of rotation curves, it is difficult
to reconcile this with velocity structure.  

3. There are dips in the continuum sources shortward of the emission edge,
particularly the lowest one in the diagram, indicative of absorption.
However, the signal is weak and it is difficult to interpolate the continuum
in the plots, as shown in Figure \ref{profiles}; therefore we can only claim
that there is a suggestion of real absorption, with low significance.

4. The [O II] image does not show a shortward edge, as expected if the
L$\alpha$ edge is due to absorption, since [O II] is a forbidden line.  
While the signal is low, the emission does have wings that lie in the
velocity region ``shortward'' of the edge in L$\alpha$. 
While quite suggestive, we note that the absence of an 
edge in [OII] could also be explained if the [OII] emitting
gas is distributed differently (in velocity or position) from that of L$\alpha$.

Given this tentative evidence, we suggest as an alternative that the
edge is a true absorption edge, due to  an expanding envelope that lies across 
the whole line of sight.
A similar phenomenon is seen in L$\alpha$ observations of $\eta$ Carinae: a
curved shifted absorption edge over the length of the extended emission
region (Gull, private communication).  While we do not suggest strong similarities, 
it may be that the cD galaxy is surrounded by an outward-moving
wind with a terminal velocity that is well-defined and about the same over
all directions covered by the line of sight. The curvature could arise
by projection effects if the expansion were at a constant velocity and
in a spherical envelope.  
 
To establish the shortward edge velocity under this assumption, we use the wavelength scale from 
the standard wavelength calibration for the STIS spectrum,
with a zero point offset for the wide slit obtained by setting the centre of
the strong and sharply-edged geocoronal L$\alpha$ emission at 1215.7\AA.
The absorption edge wavelength was measured from plots of sections in the spectral
image, as the 10\% point up the short wavelength side of the emission
(see examples in Figure~\ref{profiles}).
The absorption edge velocities were mapped as far as emission is
detected in both directions away from the nucleus.

These edge measurements are shown in Figure~\ref{vel}, referred to as velocities 
with respect to the overall galaxy redshifted wavelength. This implies an
expansion velocity of about 5000 km s$^{-1}$. If we fit a simple
expanding spherical absorber model to the curve, 
moving outwards at this velocity, its radius is about 12 Kpc.
This is not very sensitive to the zero point of the expansion: at
6000 km s$^{-1}$, the implied radius is 13 Kpc.  However, if the emission arises
spatially along the edge of the slit, we may have velocities as low as
0 km s$^{-1}$, with respect to the galaxy mean redshift.
The absorption trough seen against the continuum regions appears to extend 
to some 10000 km s$^{-1}$, but its outer limit cannot be determined very well
since the trough becomes weaker as it merges with the background or the
weak UV continuum where present (see Figure~\ref{profiles}). 

\section{Discussion}\label{sec-discuss}
There is probably
a mixture of distributed emission in this galaxy, including diffuse emission 
spatially slanted relative to the slit, and resolved
emission from the bright knots.  We require narrower slit observations
to untangle the situation; also, it would be of great interest to
obtain undispersed images in H$\alpha$ and [O~II] or [O~III] at HST resolution
to map the gas in the required detail.

We note that the shortward edge in L$\alpha$ is not centred on what we regard
as the nucleus, but some 0\farcs6 below it in our diagrams. The edge wavelength
(outflow velocity?) is most extreme next to the brightest UV source, so that 
possibly this represents
the central force driving any outflow. This knot accounts for 20\% of the
L$\alpha$ flux and has magnitude close to 22.3 in both filters (V and I),
corresponding to absolute magnitude -18 (similar to the luminosity of the SMC).
If this emission is due to photoionisation by massive stars, it is of interest to
see if the implied star formation rate can provide enough energy to drive the
purported outflow.  From the H$\alpha$ photometry, we use the relation of 
Kennicutt (1998) to estimate star formation rates  (assuming an
extinction of 1 mag at H$\alpha$ and an [NII]/H$\alpha$
ratio of 0.5 (Kennicutt 1992)).  This results in a star formation rate 
for the nucleus of 28 $\pm 5$ $M_\odot \mbox{yr}^{-1}$,  
where the uncertainty reflects the estimated zero point
uncertainty.  Star formation rates of the same order of magnitude are measured
in the two other resolved knots.  We calculate the total energy released by
supernovae, assuming a frequency of 0.005 per solar mass formed, a total energy per event of $10^{51}$ ergs, 
and that 10\% of this energy is transfered to the gas.  
Thus, a star formation
rate of 30 $M_\odot \mbox{yr}^{-1}$ corresponds to an energy release of 1.5$\times 10^{49}$
ergs yr$^{-1}$; this amounts to copious amounts of energy if the burst lasts several 100 Myr.  
However, this energy must go into not only the kinetic energy of the gas, but to 
overcoming the cD gravitational potential and the pressure of the intracluster medium (ICM).  
It turns out that the latter is the most
significant.  Assuming an isothermal gas density profile, the gas pressure is given by:
\begin{equation}
P={1 \over 2\pi G} \left({kT \over \mu m_H}\right)^2{\Omega_b \over \Omega_\circ} r^{-2},
\end{equation}
where kT=9.5 keV is the gas temperature (David et al. 1993), $\mu=0.59$ is the mean
molecular weight of the gas, $\Omega_b=0.0125 h^{-2}$ is the baryon density parameter
predicted by nucleosynthesis (Copi, Schramm \& Turner 1995), the total matter
density $\Omega_\circ$=0.3, and $r$ is the distance from the centre of the cluster.
The total resistant force on an expanding sphere of material ($4\pi r^2 P$) is therefore independent
of distance.  The work needed to push a shell of material out to $r=12$ kpc
against this pressure is $1.0 h^{-2} \times 10^{61}$ ergs; this is much more energy than
is available from star formation alone (even if we consider additional energy from stellar winds).  
If the expanding shell is a real structure, there must
be another source of energy driving it, or the ICM pressure must be much lower than we
have estimated (as will be the case if, for example, the gas density is much lower than that of an isothermal sphere).

The origin of the blue lane to the SE of the nucleus is uncertain.   
If the L$\alpha$ detected to the upper left in Figure \ref{wfpc} is associated with this lane (as we 
assume for the solid line in Figure \ref{vel}), this gas is approaching at
very high velocity.  However, H$\alpha$ is
only strongly detected to the NW, where the gas velocity is clearly receding.  It would be
unusual to detect only the receding side of a double jet system in H$\alpha$.
It seems likely, then, that the blue knots and nebular emission arising from the
NW are due to infalling nebulae, and that the L$\alpha$ to the SW is {\it not}
associated with the blue lane to the SE.  It is more likely that the L$\alpha$ traces
the faint H$\alpha$ SW extension, and that the linear, blue feature is not associated with
any of the emission features.  However, narrower slit observations are required to
establish this with certainty.  One possible explanation is that the blue lane corresponds
to a conical hole blown through the intergalactic medium by an active nucleus; if this
hole is devoid of dust, and/or lined with hot stars, it may appear blue, but without nebular
emission lines. 

\section{Summary}\label{sec-summary}
   Our data show that there is extended line emission in L$\alpha$, H$\alpha$, and [OII] 
in the cD galaxy at the centre of Abell 2390.
This emission is seen
distributed non-uniformly across and around the galaxy, in both diffuse and knotty 
components.  It is difficult to separate the spatial and velocity shifts in our dispersed
images, due to the wide slit; however, there is strong evidence for high velocities due
to infalling nebulae.  There is a sharp edge in the L$\alpha$ spectrum which, if due to
absorption, may indicate outflow from the whole galaxy with very high
velocity.   It is unlikely that the star formation rates derived from the H$\alpha$ flux
produce enough energy (through supernovae) to overcome the ICM pressure and produce
such an outflow.
   
If the detected line emission arises from massive star formation, this implies 
that there is ongoing evolution of the cD galaxy.
Young populations are found only in a few CNOC1 clusters by Davidge and Grinder (1996), 
and are rare in low redshift rich clusters (e.g. Oegerle and Hoessel 1991).
These data provide additional evidence that clusters with strong cooling flows
detected in X-rays have strong emission lines, with asymmetric distributions
and spatially varying line ratios (e.g. Crawford et al. 1999).  We suggest that
the difference between the H$\alpha$ and L$\alpha$ images is the result of
inhomogeneous dust extinction.

More detailed spectral
mapping in other lines with high spatial resolution would be valuable in
determining the precise velocity structure of this system.
It would also be valuable to extend the sample of such observations to
include cD galaxies with a wide range of line intensity, to quantify 
the connection between star-formation, inflows/outflows, and cooling
flows.  In particular, it would be of interest to study 
the other two CNOC1 cD galaxies with similarities to that of A2390, 0839+29 and 1455+22 
(Davidge \& Grinder 1996), in similar
detail.

  We thank Tim Davidge for helpful discussions on the results in
this paper, and Alastair Edge for a careful reading of the manuscript and
many useful suggestions.  
When this work was begun, MLB was supported by a Natural Sciences and Engineering 
Research Council of Canada (NSERC)
research grant to C. J. Pritchet and an NSERC postgraduate scholarship.  MLB is currently
supported by a PPARC rolling grant for extragalactic astronomy and cosmology at Durham.
  
\newpage   

\begin{deluxetable}{llll}
\tablenum{1}
\tablecaption{Observations of central A2390}
\tablecaption{Slit 52 x 2 arcsec, PA 154$^o$}
\tablehead{\colhead{HST \#} &\colhead{Grating} &\colhead{Exposure} 
&\colhead{Comment}}
\startdata
O4VY01070 &G430L &1500sec &CCD\nl
O4VY01080 &G750L &150sec &CCD, v weak\nl
O4VY010B0 &G140L &2300 sec &FUV MAMA\nl
O4VY010C0 &G140L &2300 sec &FUV MAMA\nl
\enddata

\end{deluxetable}

\newpage

\centerline{Captions to Figures}

\figcaption[wfpc.ps]{Four aligned images of the cD galaxy in A2390.
>From left to right: (1) H$\alpha$ CFHT image; (2) F555W from WFPC2; 
(3) ratio F814W/F555W (light=blue); (4) The dispersed image
of L$\alpha$ matched in spatial scale.  The solid, horizontal line
in the second panel is 2\arcsec\ long, and shows the 
width and placement of the slit; the position angle is 154$^o$.\label{wfpc}}

\figcaption[uvspec.ps]{The UV spectral image of the cD galaxy of A2390 slightly
smoothed and compressed by a factor of 2 in the spatial direction. The 
wavelength coverage is roughly 1330\AA\ to 1650\AA. Note the three continuum sources and
the curved absorption edge to the L$\alpha$ emission structure. \label{uvspec}}

\figcaption[profiles.ps]{Line profiles from STIS spectral images. Note the
sharp shortward cutoff seen in L$\alpha$ but not [O II]. The L$\alpha$
knot has a continuum, sketched in roughly, possibly indicating absorption out to
velocities of about -10000 km s$^{-1}$. All velocities refer to the galaxy 
mean redshift of 0.2301, and errors are 100 km s$^{-1}$ or less. \label{profiles}}

\figcaption[vel.ps]{Velocities of measured L$\alpha$ features, relative to the
cD galaxy.  The connected 
short dashes are emission features spatially identified with the blue continuum peaks. 
If, instead,  the L$\alpha$
emission traces the H$\alpha$ emission, the velocities are given 
by the dots. Vertical lines represent the locations of detected UV continuua. 
The dashed curve is the inner L$\alpha$ edge velocity, assuming that it is an absorption feature 
and that the emission arises in the galaxy center or knots. 
\label{vel}}

\newpage

\centerline{References}

Abraham R.G., \it et al \rm, 1996, ApJ, 471, 694

Allen S. W. 1995, MNRAS, 276, 947

Allen S. W., Fabian A. C., Edge A. C., Bohringer H., White D. A. 1995, MNRAS, 275, 741


Balogh, M. L. 1999, Ph.D. thesis, University of Victoria

Balogh, M. L., Morris, S. L. 1999, in preparation

Conselice, C. J., Gallagher, J. S., Calzetti, D., Homeier, N., Kinney A. 1999, ApJ, accepted (Astro-ph-9910382)

Copi C. J., Schramm D. N., Turner M. S. 1995, {\it Science}, 267, 192

Cowie L. L., Hu E. M., Jenkins E. B., York, D. G. 1983, ApJ, 272, 29

Crawford C. S., Allen, S. W., Ebeling, H., Edge, A. C., Fabian, A. C. 1999, MNRAS, 306, 857

Crawford C. S., Fabian A. C. 1992, MNRAS, 259, 265

David L. P., Slyz A., Jones C., Forman W., Vrtilek S. D., Arnaud K. A. 1993, ApJ, 412, 479

Davidge T.J. and Grinder M., 1995, AJ, 109, 1433

Draine B. T. \& Salpeter E. E. 1979, ApJ, 231, 77


Edge A. C., Ivison, R. J., Smail, I., Blain, A. W., Kneib, J.-P. 1999, MNRAS 306, 599

Filippenko A. V., Terlevich R., 1992, ApJ, 397, L79

Haschick A. D., Crane P. C., van der Hulst, 1982, ApJ, 262, 81

Heckman T. M., Baum S. A., van Breugel W. J. M., McCarthy P., 1989, ApJ, 338, 48

Hu E. M. 1992, ApJ, 391, 608 

Hutchings J.B. \it et al \rm 1998, ApJ, 492, L115

Johnstone R. M., Fabian A. C., Nulsen P. E. J. 1987, MNRAS, 224, 75

Kennicutt, R. C. 1992, ApJ, 388, 310

Kennicutt, R. C. 1998, ARA\&A 36, 189

Lemonon, L., \it et al \rm 1998, A\&A, 334, L21

McNamara B.R., O'Connell R. W., 1993, AJ, 105m 417

McNamara B. R., Wise M., Sarazin C. L., Jannuzi B. T., Elston R., 1996, ApJ, 466, 66

Oegerle W.R., and Hoessel J.G., 1991, ApJ, 375, 15


Pierre M., Le Borgne, J.-F., Soucail, G., Kneib, J.-P. 1996, A\&A, 311, 413

Pinkney J., et al. 1996, ApJ, 468, L13

Romanishin W., Hintzen P., 1988, ApJ, 227, 131

Smail I., Edge A. C., Ellis R. S., Blandford R., 1998, MNRAS, 293, 124


Yee H.K.C., \it et al \rm 1996, ApJS, 102, 289

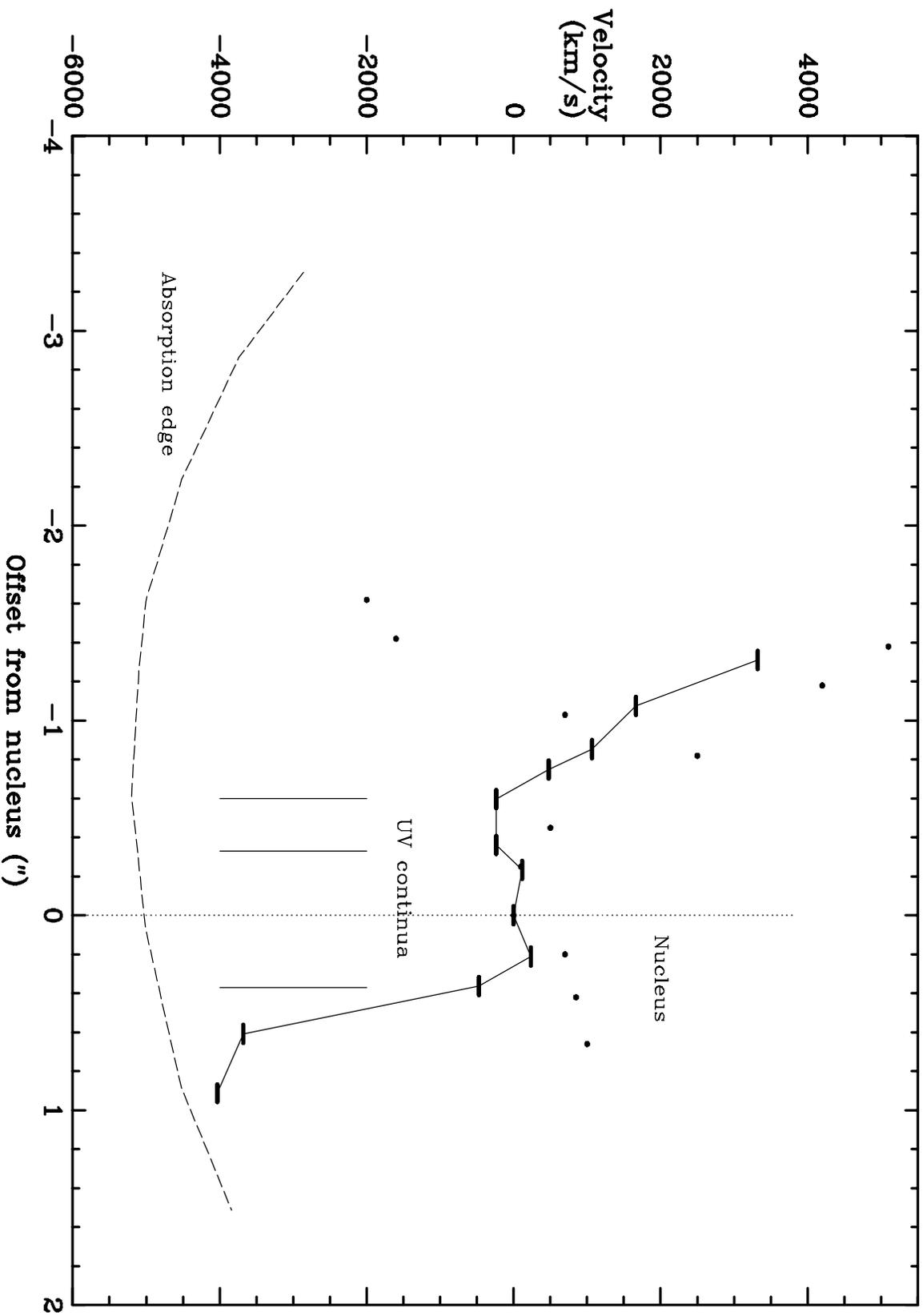
\end{document}